\documentclass[aps,english,prb,twocolumn,showpacs,superscriptaddress]{revtex4}
\usepackage{graphicx}
\makeatletter
\def\p@figure{Figure~}
\makeatother

\begin{document}

\title{Nanoscale Charge Balancing Mechanism in Alkali Substituted Calcium-Silicate-Hydrate Gels}

\author{V. Ongun \"{O}z\c{c}elik}\email{ongun@princeton.edu}
\affiliation{Andlinger Center for Energy and the Environment, Princeton University, New Jersey 08544 USA}
\affiliation{Department of Civil and Environmental Engineering, Princeton University, New Jersey 08544 USA}
\author{Claire E. White}\email{whitece@princeton.edu}
\affiliation{Andlinger Center for Energy and the Environment, Princeton University, New Jersey 08544 USA}
\affiliation{Department of Civil and Environmental Engineering, Princeton University, New Jersey 08544 USA}

\begin{abstract}

Alkali-activated materials and related alternative cementitious systems are sustainable material technologies that have the potential to substantially lower CO$_2$ emissions associated with the construction industry. However, the impact of augmenting the chemical composition of the material on the main binder phase, calcium-silicate-hydrate gel, is far from understood, particularly since this binder phase is disordered at the nanoscale. Here, we reveal the presence of a charge balancing mechanism at the molecular level, which leads to stable structures when alkalis (i.e., Na or K) are incorporated into a calcium-silicate-hydrate gel, as modeled using crystalline 14{\AA} tobermorite. These alkali containing charge balanced structures possess superior mechanical properties compared to their charge unbalanced counterparts. Our results, which are based on first-principles simulations using density functional theory, include the impact of charge balancing on the optimized geometries of the new model phases, formation energies, local bonding environments, bulk moduli and diffusion barriers of the alkali atoms within the crystals.

\end{abstract}

\maketitle

Material technologies that possess lower environmental footprints have gained significant focus over recent decades which initiated the efforts of replacing conventinal materials with their sustainable counterparts. Ordinary Portland cement (OPC) concrete, which is the most widely available human-made material,  has been used for centuries as a construction material with wide ranging applications from bridge decks to nuclear waste encapsulation while the production of OPC clinker (an ingredient required to make OPC concrete) creates up to 8\% of human-made CO$_2$ emissions and this is estimated to grow rapidly in coming years.\cite{scrivener2008innovation} Recently, alkali-activated materials (AAMs) have been proposed as an alternative to conventional OPC and are gaining interest due to the fact that AAMs can reduce CO$_2$ emissions by up to 80\% relative to OPC.\cite{l2016alkali,myers2015role,myers2013generalized} Despite the extensive history of OPC concrete usage throughout the world, there are still ambiguities regarding the atomic structure of its most critical phase, calcium-silicate-hydrate (C-S-H) gel, which is responsible for the strength and durability of cement paste.\cite{allen2007composition, richardson1999nature, qomi2014combinatorial,skinner2010nanostructure} The alkali activation process further increases the complexity of the atomic arrangements in C-S-H gels because of the increased heterogeneity of their chemical and physical nature. Experimental studies show that the inclusion of alkali atoms (Na/K) and Al in the C-S-H gel (forming C-N-A-S-H gel, where N represents the alkali and aluminum is shown by A) has critical effects on the physical and chemical properties of the structure such as increasing the OH$^-$ concentration,  changing the H$_2$O content, and causing a decrease in the basal spacing of the unit-cell and silica mean chain length.\cite{l2016alkali,l2016influence,faucon1999aluminum} It is generally accepted that the inclusion of alkali atoms alters the local bonding environment of C-S-H gels, however a comprehensive understanding of the atomistic mechanisms underlying the formation of these new materials is missing and the impact of these alkali atoms on the relative stability of the phases is unknown.

\begin{figure}
\includegraphics[width=8cm]{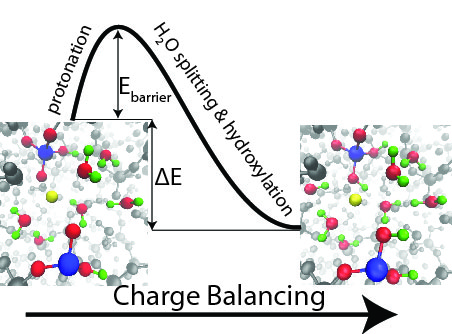}
\caption{Schematic illustration of the charge balancing process which consists of three stages: protonation by the addition of H atoms, H$_2$O splitting and hydroxylation.}
\label{figtoc}
\end{figure}

In this letter, we present a fundamental charge balancing mechanism which reveals the formation energies of a range of atomic structures representative of those found in OPC and AAM systems as well as uncovering the impact of alkali substitution on the stability of C-S-H gel. Briefly, this charge balancing is initiated by the diffusion of H atoms to the crystal structures in order to counter balance the water deprotonation that occurs as a result of substituting the intra-layer Ca atoms by alkali atoms. The additional H atoms modify the chemical bonds of the structure and initiate chain reactions within the unit-cell which lead to new structural forms of model phases representing Ca-rich AAM pastes. These new phases show similar structural stabilities and bond strengths at the molecular level as compared to C-S-H gel (represented by 14{\AA} tobermorite in this letter). Depending on the concentration and type of alkali atoms used for activation, the intermolecular bonding strengths, bulk moduli, unit-cell volumes and water contents of these phases can be modified. Additionally, substituting one of the bridging Si atoms with an Al atom creates C-A-S-H gels whose atomic arrangements are also modified with the same charge balancing mechanism, resulting in structures with higher levels of phase stability. Given the fact that AAMs are crucial materials for reducing human-made CO$_2$ emissions, we believe that (i) the fundamental charge balancing mechanism proposed in this letter, (ii) the atomistic properties associated with this mechanism and (iii) the new model structures presented here will pave the way to a deeper understanding of the impact of alkali activation on the resulting gel phase and enhance the uptake of environmentally friendly construction materials.

\begin{figure}
\includegraphics[width=9cm]{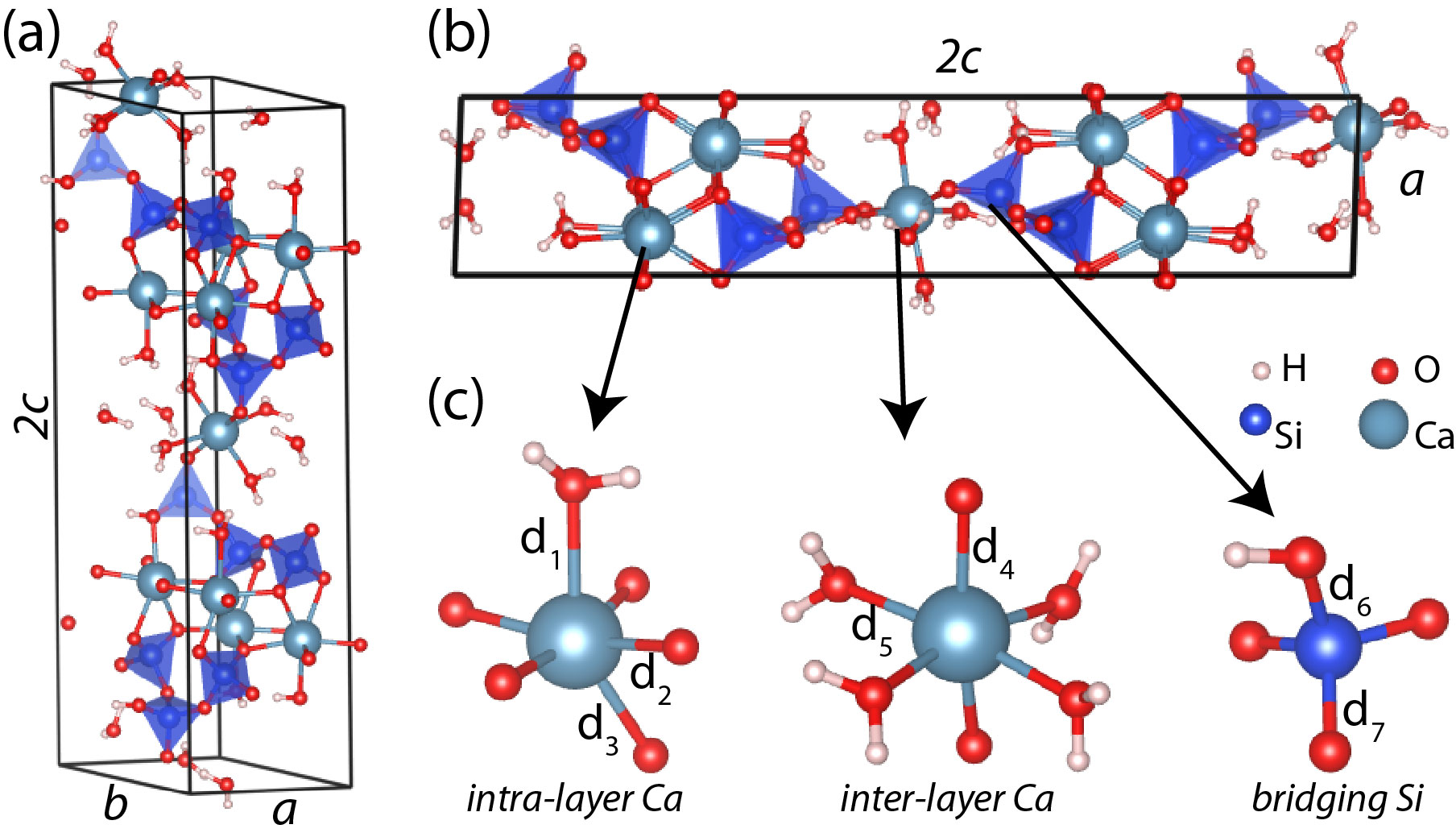}
\caption{Model C-S-H structure represented by  14{\AA} tobermorite in (a) perspective and (b) side view. (c) Bonding geometries of intra-layer Ca, inter-layer Ca and bridging Si. O, H, Si and Ca atoms are shown with red, white, blue and light blue spheres using a ball-and-stick model. Selected bond lengths and unit-cell dimensions are indicated where d$_1$=2.57{\AA}, d$_2$=2.38{\AA}, d$_3$=2.42{\AA}, d$_4$=2.30{\AA}, d$_5$=2.45{\AA}, d$_6$=1.67{\AA}, d$_7$=1.60{\AA}, a=6.57{\AA}, b=7.36{\AA}, 2c=28.01{\AA} }
\label{fig1}
\end{figure}

\begin{figure*}
\includegraphics[width=12cm]{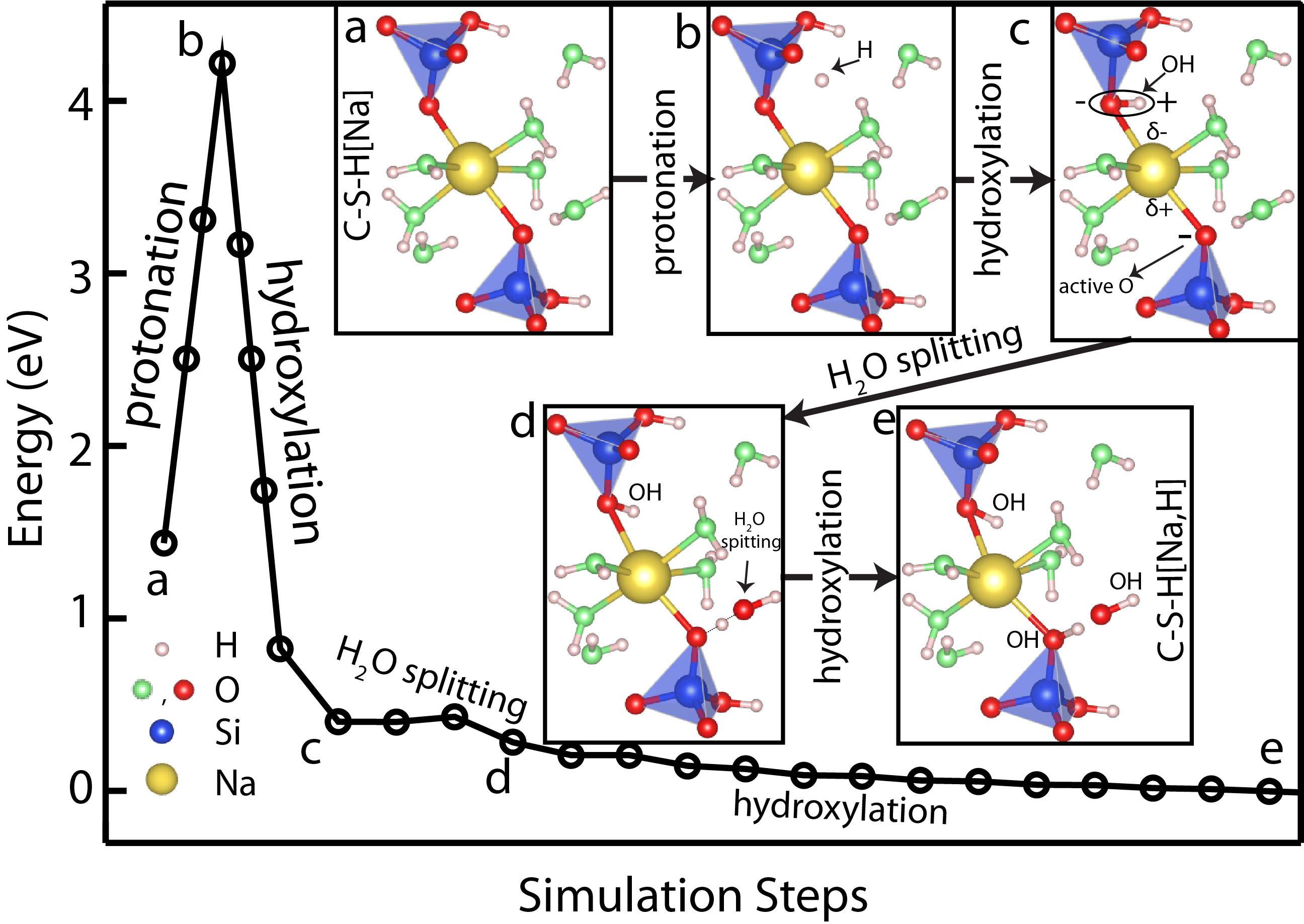}
\caption{The variation of total energy during the charge balancing process of the C-S-H model structure (based on 14{\AA} tobermorite) where the energy of the optimized charge balanced structure is taken as zero.  The insets (a-e) show the self-rearrangements of atoms during charge balancing. (a) Optimized Na-doped C-S-H[Na]. (b) Protonation by addition of H to the inter-layer region. (c) OH formation and charge induction on the alkali atom. (d) Charge balancing via dissociation of H$_2$O. (e) Formation of additional OH. In the ball-and-stick model, oxygen atoms in H$_2$O and OH are shown with green and red spheres, respectively. }
\label{fig2}
\end{figure*}

Although there are numerous studies on the atomic structures of C-S-H gel and tobermorite minerals,\cite{richardson2014model, myers2014thermodynamic, qomi2014combinatorial, skinner2010nanostructure,pellenq2009realistic,merlino2001real, rejmak201329si} a model derived from purely first-principles calculations without any empirical inputs is crucial in order to obtain fundamental properties such as ground state energy, internal pressure, forces on atoms, local bonding environments and distribution of local charge between the atoms. The accuracy of first-principles calculations at predicting these fundamental properties of similar cementitious materials have been justified in previous studies. \cite{saritas2015predicting, durgun2012understanding, manzano2011impact, durgun2014characterization, white2014uncovering}. Here, we focus on 14{\AA} tobermorite since this crystal structure has been used to describe C-S-H(I) gels with Ca/Si ratios less than ~1.4 and slag-based AAMs (containing C-N-A-S-H gel) tend to have Ca/(Si+Al) ratios of ~1.\cite{bonaccorsi2005crystal,richardson2013importance,myers2015role}  After optimizing the shape of the crystal unit-cell, bond lengths, orientation of H$_2$O molecules with respect to silica chains and the inter-layer spacing; we end up with the optimum ground state of 14{\AA} tobermorite with the chemical formula of Ca$_5$Si$_6$O$_{16}$(OH)$_2 \cdot$ 7H$_2$O as presented in ~\ref{fig1}. Using this optimized pure structure, we produce representative phases of Ca-rich AAMs by introducing alkali (Na, K) or alkali-earth (Mg) atoms to the C-S-H unit-cell. (Note that Mg is included in this study for completeness since a small amount of Mg can be found in blast furnace slag, which is used as an AAM precursor). These atoms replace the inter-layer Ca atom and after each substitution we repeat the same rigorous optimization procedure as we did for the pure case. Thus, we end up with a set of structures whose ground state properties are calculated using the same theoretical approach, which is necessary for making appropriate comparisons of stability, energetics and mechanical strength.

Once a Ca atom is replaced by an alkali atom, the charge neutrality of the structure is broken in the vicinity of the substitution site, due to the fact that we are effectively removing a $^+2$ ion (Ca$^{2+}$) and replacing it with a $^+1$ ion (Na$^+$ or K$^+$). Thus, the crystal is deprotonated by creating a hole in its electronic structure which is a potential source of local instability under an external stimulus or in the presence of foreign reacting agents. Since our prime objective is to maintain structural and chemical stability of the material, we counter balance this deprotonation by introducing external H atoms and re-optimizing the atomic positions in the unit-cell to obtain charge balanced C-S-H[N,n] or C-A-S-H[N,n] structures where N represents the atom that substitutes the inter-layer Ca atom and n is the number of H atoms used for charge balancing. It should be noted that the addition of an extra H atom to the system is representative of the charge balancing mechanism that occurs during the alkali activation process when Na/K is included in the C-S-H and/or C-A-S-H gel. Hence, we are able to unravel the impact of the individual chemical steps involved in the charge balancing mechanism at the atomic level.

The variation of total energy during structural optimization is a powerful tool for investigating the chemical interactions taking place during the structural rearrangements of atoms. It also provides information about the intermediate chemical phases of optimization as well as the energy barriers that need to be overcome for the completion of these chemical reaction and formation of new phases. \cite{ozcelik2014new, gurel2014dissociative, ozccelik2014stable,  li2016markedly, jiang2016modeling, ersan2016effect} Thus, we present the variation of the total energy of C-S-H[Na,H] unit-cell during the charge balancing process and the related structural self-rearrangements of atoms in ~\ref{fig2}, where ground state energy of the charge balanced C-S-H[Na,H] structure is set to zero. Accordingly, there exist three fundamental steps (protonation by the addition of H atom, H$_2$O splitting and hydroxylation)  all of which contribute to the formation of a symmetrical charge distribution around the alkali atom. We start with the optimized structure of charge unbalanced C-S-H[Na] and the H atom is randomly penetrated in its inter-layer region (illustrated in insets a-b of ~\ref{fig2}). Due to the random initial location of the H atom, the total energy of the system suddenly increases by 2.8 eV  as the structure evolves from a to b in ~\ref{fig2}. In fact, this is the energy barrier that the H atom needs to overcome to reach its most favorable location, which is near the O atom between the bridging silica site and the alkali atom (inset c of ~\ref{fig2}). Eventually, the H atom forms an OH bond at one end of the alkali atom and the total energy of the unit-cell drops by 3.8 eV from point b to c. The newly created OH dipole momentarily polarizes the alkali atom  which then induces a negative charge of 0.8e on the O atom located at the opposite side of the alkali atom. Thus, the O atom between the lower bridging silica site and the alkali atom becomes chemically active. Due to its increased chemical activity, the free H$_2$O molecule in the inter-layer region is attracted to this O atom, (as shown by inset d of ~\ref{fig2}). Eventually, H$_2$O dissociates into a H$^+$ and OH$^-$ ions, and the H$^+$ ion establishes another OH bond near the lower bridging silicon site as shown in inset e. As a result of this sequence of chemical reactions, the addition of the initial H atom results in the formation of three extra OH groups and dissociation of one H$_2$O molecule. It should be noted that, during this process, other H$_2$O molecules bonded to the alkali atom rotate to maintain a symmetrical charge distribution. In the final optimized structure, the local charges on the atoms around the alkali atom (Na) maintain proper charge balancing with a total charge of zero. Also, as a result of this process, the total energy of the unit-cell drops by 1.4eV, resulting in a chemically stable structure. The same charge balancing steps are also observed when the unit-cell is activated by K instead of Na.

\begin{figure}
\includegraphics[width=8cm]{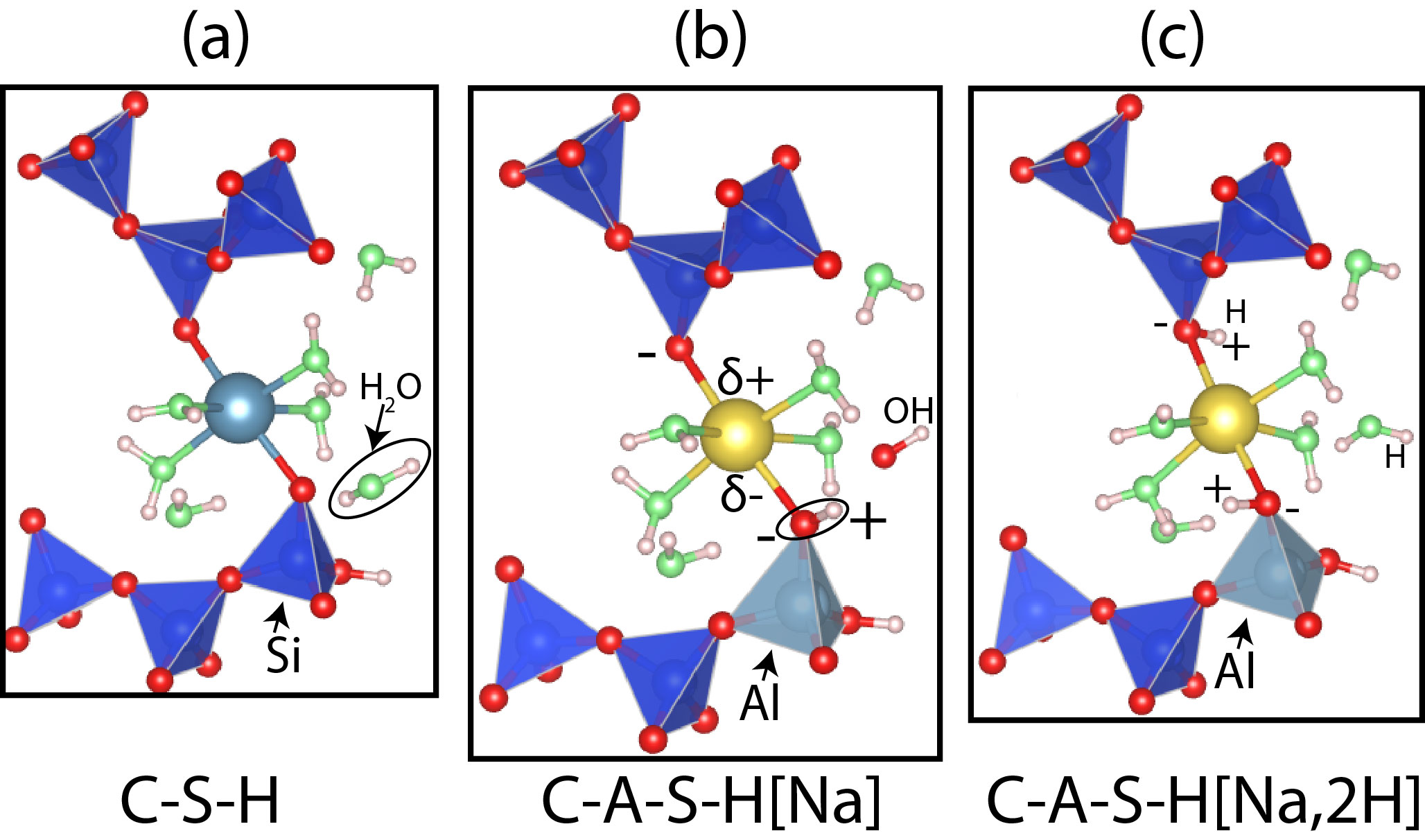}
\caption{Formation of charge balanced C-A-S-H[Na,2H] from C-S-H. (a) Initial C-S-H structure. (b) Charge unbalanced C-A-S-H structure after the substitution of the inter-layer Ca with Na and the bridging Si with Al. One water molecule is dissociated and two OH molecules are created which hydroxylate the inter-layer region (c) Charge balance is re-obtained by the addition of two external H atoms. The final structure has the same water content with the initial C-S-H structure while the inter-layer region is hydroxylated.}
\label{fig3}
\end{figure}

An important consideration for OPC concrete containing Al-rich supplementary cementitious materials (such as metakaolin) and Ca-rich AAM phases (which also contain a sizable amount of Al) is the impact of Al on the energetics and mechanical properties of the resulting binding gel. In Ca-rich AAMs, these alkali containing gels, named as C-N-A-S-H gels, are obtained by substituting a Si atom with an Al atom as well as replacing an inter-layer Ca with an alkali atom. Thus, effectively 2 protons are removed from the vicinity of the substitution site, which necessitates the addition of 2 protons to the unit-cell for proper charge balancing and chemical stability. The mechanism proposed for the alkali containing C-S-H gel is also valid for C-N-A-S-H. Here, replacing the bridging Si with Al causes this site to become negatively charged (due to the loss of a proton) which induces a negative charge on the neighboring O atom, as illustrated in ~\ref{fig3}(a-b). The negatively charged O atom attracts the nearby H$_2$O molecule and subsequently dissociates it into OH- and H+ ions. This H+ ion forms an OH molecule on the bridging site. Thus, when Si is replaced with Al, initially the unit-cell dehydrates by one water molecule and two new OH groups are created in the inter-layer region (one OH is bonded to Al and the other one is free). The OH group which is bonded to Al triggers a charge induction chain reaction in a similar fashion to the one explained in the previous paragraph. As shown in ~\ref{fig3}(c), the first externally added H atom attaches to the O atom that is at the other end of the alkali atom (and associated with the silica bridging site). The second H atom added to the system neutralizes the free OH ion present in the inter-layer, re-establishing the H$_2$O molecule that disassociated in the previous step. In the end, we acquire a charge balanced alkali containing C-A-S-H gel, (i.e., C-A-S-H[Na,2H]), that preserves the initial water content and is chemically stable. Interestingly, this process causes the protonation of the bridging sites in the (alumina)silicate chains by two new OH groups per unit-cell, which requires experimental verification. It should also be noted that the incorporation of alkali within the C-A-S-H gel reduces its unit-cell volume and basal spacing by 0.5 \%.

\begin{figure}
\includegraphics[width=9cm]{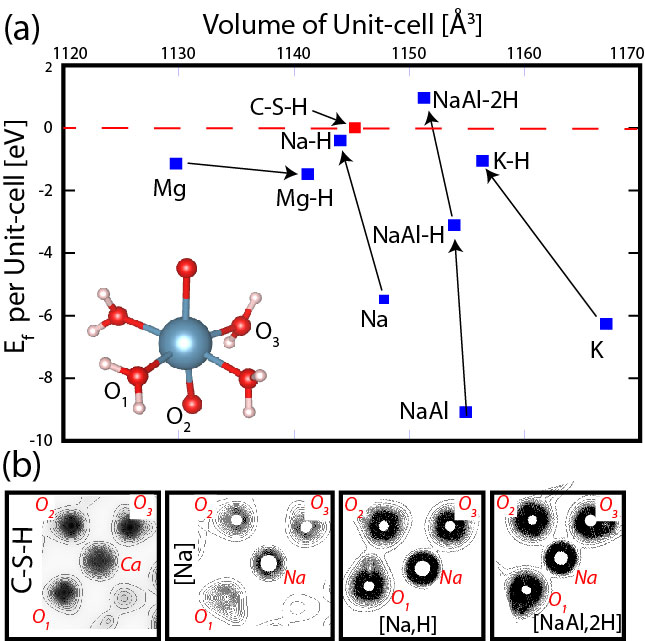}
\caption{(a) Formation energies of Na, K, Mg, NaAl doped C-S-H gel and their charge balanced structures (indicated with H or 2H) with respect to the pure C-S-H structure. The cohesive energy of pure C-S-H is taken as zero. Negative formation energies correspond to less favorable structures as compared to C-S-H. (b) Charge density contour plots on the plane passing from the inter-layer atom and its neighboring O atoms. Note that bond strength increases as H atoms are introduced for charge balancing. Atomic stuctures of the charge balanced gels are provided as Supporting Information}
\label{fig4}
\end{figure}

The thermodynamic stabilities of the charge balanced structures presented so far can be evaluated by comparing their cohesive energies. Cohesive energy per unit-cell defined by $E_{c}=\sum_{i}nE_{i}- E_{structure}$,\cite{kitel1971introduction} is the difference between the sum of energies of free atoms present in the unit-cell and the total energy of the overall structure in its optimized geometry. Here, $n_i$ is the number of $i$ atoms present in the unit-cell and $E_i$ is the ground state energy of the $i^{th}$ atom calculated in its own magnetic state. $E_c$ by itself signifies the energy gained per unit-cell by constructing a particular phase from its constituent atoms. Due to the strong covalent and ionic bonds present in these gels, all of them have high cohesive energies, indicating stability at ambient conditions. However, for practical purposes and bench marking, we need to determine the formation energies of these phases with respect to the well-known C-S-H gel which is the difference between their cohesive energies, $E_{f}[X] = E_{c}[X] - E_{c}[C-S-H]$, where X is phase of interest and  $E_{c}[C-S-H]$ is the cohesive energy of 14{\AA} tobermorite that is used to represent C-S-H gel in this study. We use these $E_{f}$ values as the primary criterion for determining whether an energy loss or an energy gain is associated with the substitution of each alkali atom such that negative values of $E_{f}$ imply less favorable structures whereas positive values of $E_{f}$ indicate more favorable structures as compared to C-S-H gel. As shown in ~\ref{fig4}(a), substitution of the inter-layer Ca with Na (or K) creates an energy loss of 5.5 eV (or 6.3 eV). The cost of K doping is slightly higher than Na doping, which is a result of the larger atomic radius of K causing greater distortion of the unit-cell. In a similar fashion, NaAl doped C-S-H gel  (or C-A-S-H[Na]) is also less favorable than the initial C-S-H gel by 9.1 eV. After the charge balancing process the energy losses of Na and K doped structures, namely C-S-H[Na,H] and C-S-H[K,H], reduce to 0.4 and 1.1 eV, respectively. Remarkably, charge balancing the Na doped C-A-S-H gel by addition of two H atoms results in the C-A-S-H[Na,2H] structure which is even more favorable then the initial C-S-H gel by 0.95 eV. Apart from improving the stabilities of the alkali containing gels, charge balancing also creates more compact and tightly bound crystal structures, as evidenced by the reduction of the crystal unit-cell volumes by up to 2\%. It should be noted that, as opposed to alkali substitution, Mg substitution creates a lower energy loss of only 1.1 eV due to the fact that Mg also has a +2 ionic configuration, similar to Ca. As expected according to charge balancing arguments, C-S-H[Mg,H] structure is a positively charged model and enlarges the unit-cell volume by 1.4 \%, resulting in a structure which is 0.15 eV less favorable than the initial C-S-H[Mg] structure.

\begin{figure*}
\includegraphics[width=12cm]{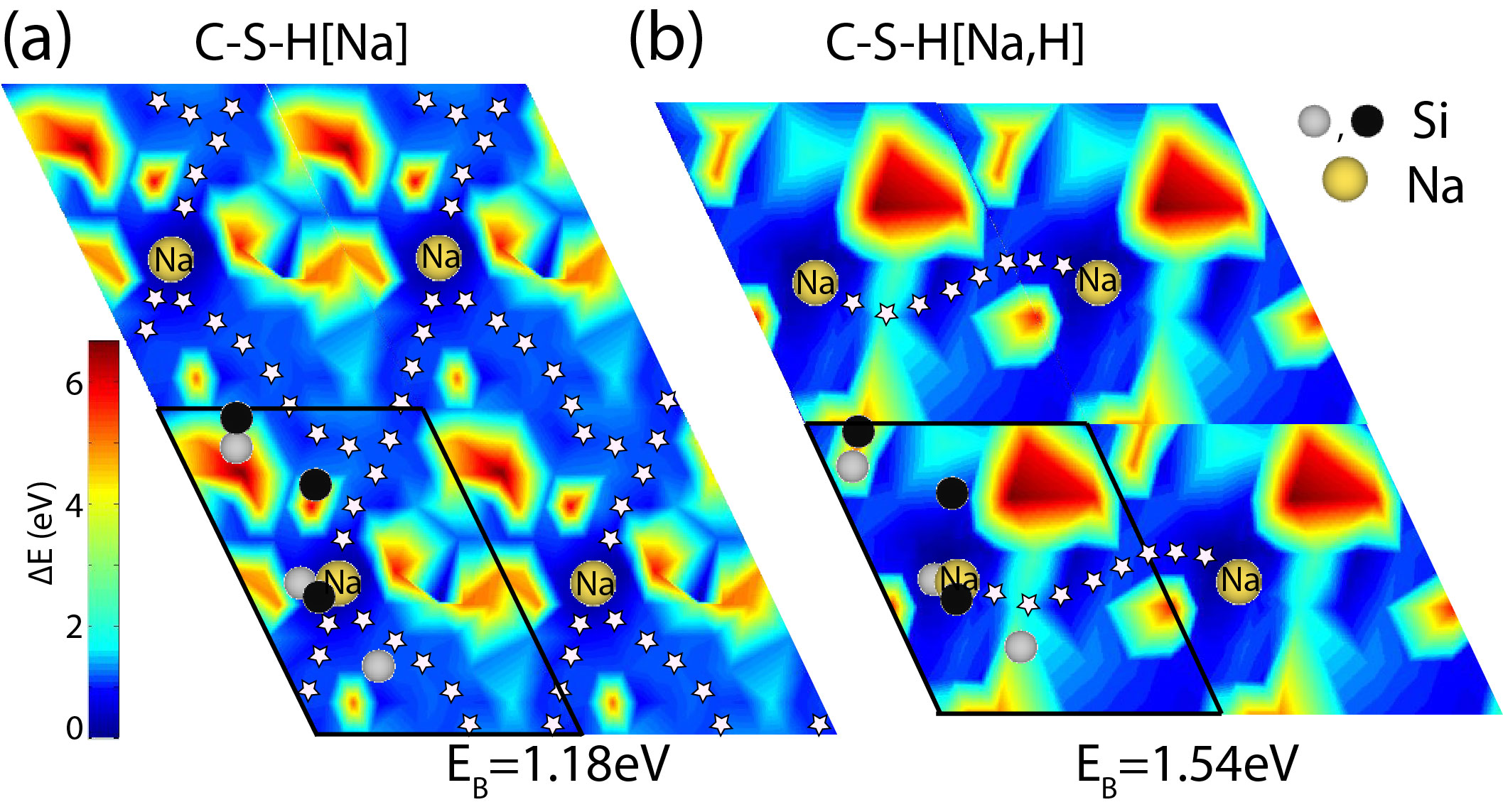}
\caption{(a)Energy landscape for the Na atom in the inter-layer region of C-S-H[Na] plotted on the 2x2 unit-cell. Its migration path along the minimum energy barrier is indicated by white stars. (b) Energy landscape for the charge balanced C-S-H[Na,H]. Note that, upon charge balancing, the activation energy for diffusion increases from 1.18 eV to 1.54 eV, which is an indication of increased mechanical strength. In the energy landscapes, blue regions represent more favorable sites for the Na atom as compared to red regions. Si atoms below and above the Na atom in the inter-layer region are represented by gray and black spheres respectively, and the diffusing Na atom is represented by yellow sphere.}
\label{fig5}
\end{figure*}

The improved stability of the crystals after charge balancing is also apparent in the local charge densities. In ~\ref{fig4}b, we present charge density contour plots for C-S-H, C-S-H[Na], C-S-H[Na,H] and C-A-S-H[Na,2H] structures, calculated on the plane passing through the inter-layer Ca/Na site and O atoms (as depicted in ~\ref{fig4}a). In each calculation, the contour interval is chosen to be the same (0.1 eV/A$^2$) and a denser grid implies stronger bonding. It is seen that the inter-layer Ca atom has stronger bonds with its neighboring O atoms in C-S-H gel as compared to weaker bonds of the inter-later Na atom in C-S-H[Na]. However, these weak bonds regain strength upon proper charge balancing of the unit-cell and eventually C-A-S-H[Na,2H] gel has the strongest local bonds around the inter-layer atom. As further confirmed by the variation of bond strengths, charge balancing of the alkali containing structures via addition of H atoms results in phases that are chemically more stable.

We finally turn our attention to the effects of charge balancing on the macroscopic observable signatures of stability such as the bulk moduli of the crystals and the diffusion barriers of atoms within the unit-cell. The bulk modulus of a crystal can be obtained using the equation $K=-V^{opt}dP/dV = -V^{opt} d^2E/dV^2$ where the second derivative of the total energy (E) with respect to unit-cell volume (V) is calculated in the vicinity of $V^{opt}$ (volume corresponding to the energetically most favorable configuration), within the harmonic approximation. While calculating the energies in the vicinity of V$^{opt}$, the volume of the unit-cell was uniformly changed within the range of $\pm$ 3 percent of V$^{opt}$ and the positions of the atoms within the unit-cell were re-optimized for each volume.  Specifically, the bulk modulus of 14{\AA} tobermorite (representing C-S-H gel) which is calculated as 54 GPa, is reduced to 49 GPa for C-S-H[Na] (36 GPa for C-S-H[K]). Balancing the local charge increases this value to 52 GPa for C-S-H[Na,H] (40 GPa for C-S-H[K,H]), which further increases to 66 GPa for the charge balanced C-A-S-H[Na,2H] gel.  These calculated bulk modulus values are in the same range with the previously reported experimentally obtained value of 47 GPa for 14{\AA} tobermorite\cite{oh2012experimental} and other computational studies which use ab-initio and force field methods.\cite{shahsavari2009first, manzano2007mechanical, manzano2009elastic, jackson2013material} In parallel with the formation energy and local bond strength results, the charge balancing process increases the bulk modulus of the alkali containing gels.

The bulk modulus of a crystal structure is closely related with the activation energy for the diffusion (or diffusion barrier) of its constituent atoms.\cite{wentzcovitch2009anomalous} Thus, as another way of demonstrating the effect of charge balancing on mechanical strength, we finally calculate the diffusion barrier of the inter-layer Na atom in the model structure of C-S-H[Na] gel and compare it with the mobility of the Na atom in the charge balanced C-S-H[Na,H] gel structure. The minimum diffusion barriers have been obtained by analyzing the energy landscapes of the Na atom in the inter-layer region, which are shown in ~\ref{fig5}. The energy landscape of each structure was calculated by manually placing the Na atom at various positions in the inter-layer region and performing self-consistent geometry optimization for the surrounding atoms. During this geometry optimization, the Na atom is kept fixed at a particular position on the horizontal $xy$ plane while its $z$ coordinate is relaxed which allows us to scan a 3D energy path for the inter-layer atom. This calculation was repeated to obtain the ground state energies for a total of 100 points that are uniformly distributed on the $xy$-plane of the unit-cell and the energy values of the remaining points were determined by interpolation of the calculated data. Thus, possible migration paths for the Na atom, which align with the minimum diffusion barriers, are obtained as indicated by white stars in ~\ref{fig5}. It is seen that the Na atom needs to overcome a barrier of 1.18 eV in the C-S-H[Na] structure, whereas this barrier increases to 1.54 eV for the the charge balanced structure  C-S-H[Na,H]. The increase in the diffusion barrier upon charge balancing provides further evidence of improved structural stability via the symmetrical charge balancing mechanism proposed in this letter. It should also be noted that, the migration for the pure C-S-H structure was calculated as 1.61 eV, and the trends in migration barriers are consistent with the trends in bulk modulus and formation energy values of these structures.

In conclusion, the proposed charge balancing mechanism and models derived using this mechanism in this letter shed light on the fundamental structural arrangements at the atomistic level that contribute to the macroscopically measured properties of cement-based gels. We show that the charge balancing process, which consists of protonation, hydroxylation and water dissociation steps, is a dominant factor governing the ground state structures of the atomistic models representative of these low-CO$_2$ materials. The inclusion of Na (or K) in the gel can lead to a highly stable phase, but only if the alkali is accompanied by a bridging Al site and the proper amount of H atoms for charge balancing. These models imply that the long term phase stability and mechanical strength of Ca-rich AAMs are in the same range of C-S-H-based OPC systems, provided that proper charge balancing takes place during their synthesis. Further research can make it possible to assess the impact of nanoscale structural disorder on phase stability, which may be an important aspect governing the mechanical and physical properties of these alkali activated gels.

\small
\section*{Methods}
Our predictions are obtained from state of the art first-principles pseudopotential calculations based on the spin-polarized DFT within generalized gradient approximation (GGA) including van der Waals corrections.\cite{grimme2006semi} We used projector-augmented wave potentials (PAW)\cite{blochl94} and the exchange-correlation potential is approximated with Perdew-Burke-Ernzerhof (PBE) functional.\cite{pbe} The Brillouin zone (BZ) was sampled in the Monkhorst-Pack scheme, where the convergence in energy as a function of the number of \textbf{k}-points was tested. The \textbf{k}-point sampling of (3$\times$3$\times$3) was found to be suitable for the BZ corresponding to the primitive unit-cell of 14{\AA} tobermorite. For larger supercells this sampling has been scaled accordingly. Atomic positions were optimized using the conjugate gradient method, where the total energy and atomic forces were minimized. The energy convergence value between two consecutive steps was chosen as $10^{-5}$ eV. A maximum force of 0.1 eV/\AA~ was allowed on each atom. Numerical calculations were carried out using the VASP software.\cite{vasp}  Mulliken charge densities were calculated by the SIESTA code\cite{siesta} starting from the pre-converged structures in VASP. A 200 Ryd mesh cut-off was chosen and the self-consistent field (SCF) calculations were performed with a mixing rate of 0.1. Core electrons were replaced by norm-conserving, nonlocal Troullier-Martins pseudopotentials.\cite{troullier1991} It should also be noted that the computations were performed using a 1x1x2 unit-cell (with dimensions a,b,2c) in order to account for the partially occupancies of one Ca atom and three H$_2$O molecules that are present in the atomic structure files available in the literature.

\bibliography{arxiv.bbl}

\end{document}